\documentclass[apjl]{emulateapj}
\usepackage{graphicx}
\usepackage{amssymb}
\usepackage{amsmath}
\usepackage{natbib}

\shorttitle{Alpha Elements in the Virgo Cluster Outskirts}
\shortauthors{A. Simionescu et al.}

\begin{document}

\title{A Uniform Contribution of Core-Collapse and Type Ia Supernovae to the Chemical Enrichment Pattern in the Outskirts of the Virgo Cluster}

\author{A. Simionescu\altaffilmark{1}, N. Werner\altaffilmark{2,3}, O. Urban\altaffilmark{2,3,4}, S. W. Allen\altaffilmark{2,3,4}, Y. Ichinohe\altaffilmark{1,5}, I. Zhuravleva\altaffilmark{2,3}}
\affil{$^1$Institute of Space and Astronautical Science (ISAS), JAXA, 3-1-1 Yoshinodai, Chuo-ku, Sagamihara, Kanagawa, 252-5210 Japan}
\affil{$^2$KIPAC, Stanford University, 452 Lomita Mall, Stanford, CA 94305, USA}
\affil{$^3$Department of Physics, Stanford University, 382 Via Pueblo Mall, Stanford, CA  94305-4060, USA}
\affil{$^4$SLAC National Accelerator Laboratory, 2575 Sand Hill Road, Menlo Park, CA 94025, USA}
\affil{$^5$Department of Physics, Graduate School of Science, University of Tokyo, 7-3-1 Hongo, Bunkyo, Tokyo 113-0033, Japan}

\begin{abstract}

We present the first measurements of the abundances of $\alpha$-elements (Mg, Si, and S) extending out to beyond the virial radius of a cluster of galaxies. Our results, based on Suzaku Key Project observations of the Virgo Cluster, show that the chemical composition of the intra-cluster medium is consistent with being constant on large scales, with a flat distribution of the Si/Fe, S/Fe, and Mg/Fe ratios as a function of radius and azimuth out to 1.4 Mpc (1.3 $r_{200}$). Chemical enrichment of the intergalactic medium due solely to core collapse supernovae (SNcc) is excluded with very high significance; instead, the measured metal abundance ratios are generally consistent with the Solar value. The uniform metal abundance ratios observed today are likely the result of an early phase of enrichment and mixing, with both SNcc and type~Ia supernovae (SN~Ia) contributing to the metal budget during the period of peak star formation activity at redshifts of 2--3. We estimate the ratio between the number of SN~Ia and the total number of supernovae enriching the intergalactic medium to be between 12--37\%, broadly consistent with the metal abundance patterns in our own Galaxy or with the SN~Ia contribution estimated for the cluster cores. 
\end{abstract}


\section{Introduction}\label{intro}

The chemical composition of the diffuse hot gas in galaxies and clusters of galaxies contains valuable information about the origin of chemical elements and their distribution during the evolution of the Universe. Elements heavier than and including oxygen are produced by supernova explosions, with core-collapse supernovae (SNcc) producing predominantly lighter metals, from O to Si and S, while type~Ia supernovae (SNIa) are responsible for heavier elements from Si to Fe and Ni \citep[e.g.][]{tsujimoto1995}. These elements are then expelled from their host galaxies by galactic winds and ram-pressure stripping. The metal abundances in the intergalactic space are therefore very sensitive to the numbers of SN~Ia and SNcc that contributed to the enrichment, as well as to the time-scale over which the supernova products were injected into the intra-cluster medium (ICM). The initial metallicity of the progenitors, the initial mass function (IMF) of the stars that explode as SNcc, and details about the exact SNIa explosion mechanism, also influence the observed abundance ratios \citep[see][for a review]{werner2008}.

Initial results from the ASCA satellite suggested a radially increasing Si/Fe ratio in the ICM, with the X-ray gas at radii beyond about 1~Mpc having a chemical composition consistent with pure SNcc yields \citep{finoguenov2000}. The paradigm that emerged from these measurements was that of an early enrichment driven predominantly by SNcc in the proto-cluster phase, which led to the SNcc products being well mixed throughout the ICM, and a later continuous enrichment by SN~Ia with longer delay times in the cD galaxy, creating a peak in SN~Ia products at the cluster centers \citep[e.g.][]{degrandi2001,bohringer2004,tamura2004}. 

This scenario has been challenged by more recent results from the XMM-Newton, Chandra, and Suzaku satellites, reporting nearly constant or even radially decreasing Si/Fe and S/Fe profiles, and only a marginal radial increase in the O/Fe ratio \citep[e.g.][and references therein]{simionescu2009b,sato2009,simionescu2010,million2011}. However, all these measurements are concentrated within the central, brighter regions of any given cluster, therefore probing only a fraction of the total volume of the ICM. 

More recently, using a detailed study of the outskirts of the Perseus Cluster with Suzaku, \citet{werner2013nat} presented measurements of the Fe abundance in the ICM in 78 independent regions beyond the cluster core and out to the virial radius (defined here as $r_{200}$, the radius within which the mean enclosed mass density of the cluster is 200 times the critical density of the universe at the cluster redshift). However, the equivalent widths of spectral lines from other elements are much weaker than Fe and therefore alpha-elements have been significantly more challenging to detect. \citet{Sasaki2014} reported nearly constant alpha-element to Fe ratios consistent with the Solar value out to 0.5 $r_{180}$ in four nearby galaxy groups, but the metal abundance ratios beyond about a half of the virial radius, as well as the chemical composition in the outskirts of more massive systems such as galaxy clusters, have not yet been measured.

Here, we present results from Suzaku Key Project observations of the Virgo Cluster, the nearest galaxy cluster at a distance of 16.1~Mpc \citep{tonry2001} and the second brightest extended extragalactic soft X-ray source. Because line emission from Si and S is maximised around the mean temperature of the Virgo ICM ($\sim2$ keV), this system presents an ideal target for determining the abundances of these metals.

\section{Observations and data reduction}

A large mosaic of Suzaku observations of the Virgo Cluster was obtained as a Suzaku Key Project during AO-7 and AO-8, totalling 60 pointings and 1~Ms of net exposure time. The observations cover the cluster out to beyond its estimated virial radius \citep[$r_{200}\sim1.1$~Mpc,][]{urban2011} along four arms extending towards the north, west, south, and east from the brightest cluster galaxy, M87. In addition, we included seven fields covering the outer parts of the cluster along the northern arm, between 1.0 and 1.5~Mpc (AO-6). 

The data from the X-ray Imaging Spectrometers (XIS) 0, 1 and 3 were analyzed following the procedure described in \citet{simionescu2013} and \citet{urban2014}. In short, in addition to applying the screening criteria recommended by the instrument team\footnote{Arida, M., XIS Data Analysis, http://heasarc.gsfc.nasa.gov/docs/suzaku/analysis/abc/node9.html (2010)}, we filtered the data to include observation periods with the geomagnetic cut-off rigidity COR $>$ 6 GV, and excluded two columns on either side of the charge-injected columns in XIS1 to avoid charge leak known to affect these areas of the detector. Furthermore, we used a mask file provided by the XIS team to manually filter out pixels that are known to be hot or flickering\footnote{http://www.astro.isas.ac.jp/suzaku/analysis/xis/nxb\_new (2015)}. We applied the latest contamination layer calibration from 2014 August 25.
 
\begin{figure}
\begin{center}
\includegraphics[width=\columnwidth]{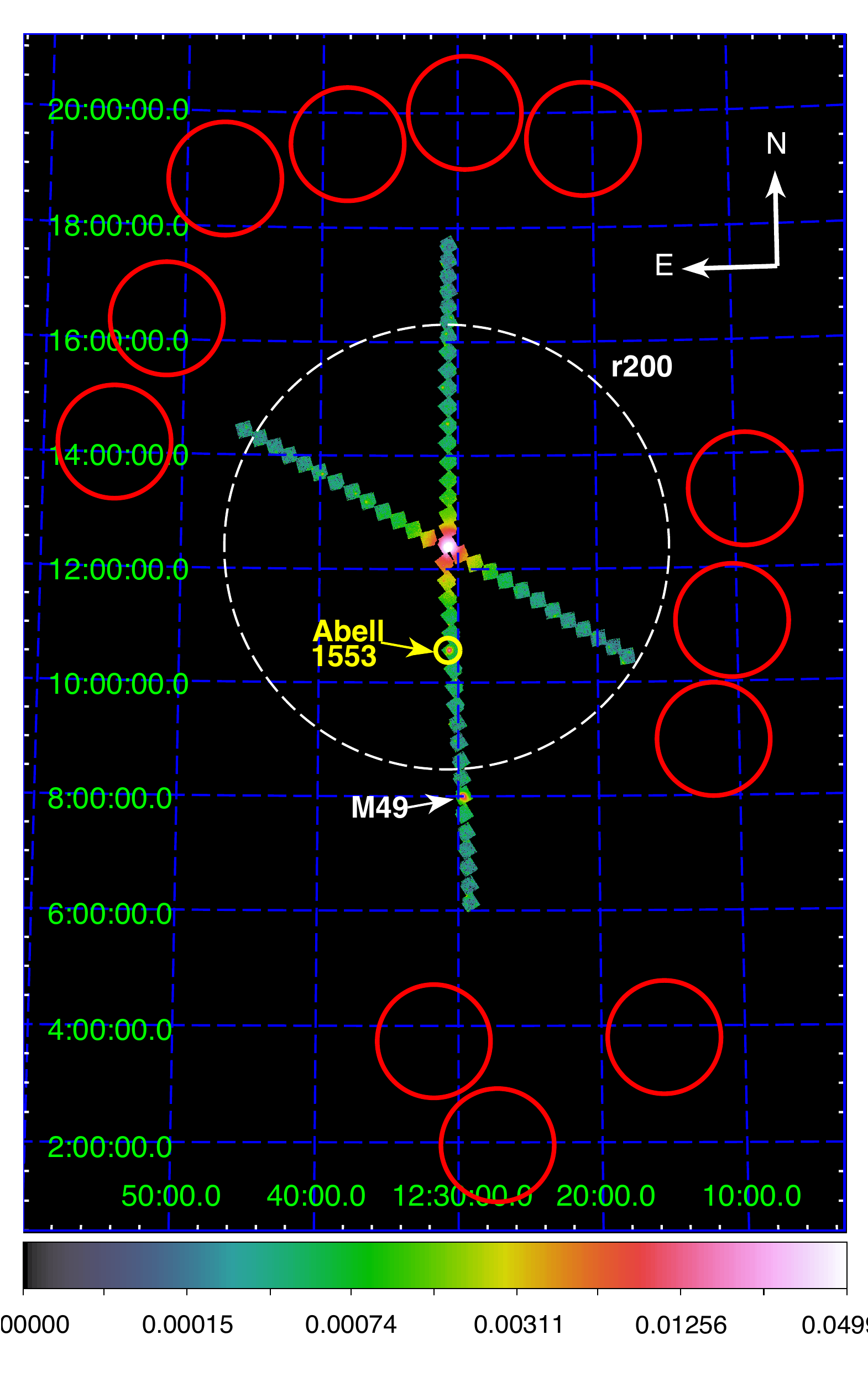}
\caption{Image of the complete Suzaku mosaic of the Virgo Cluster in the 0.7-2.0 keV energy band, corrected for instrumental background and vignetting. The white dashed circle marks the approximate location of $r_{200}$, while the red circles mark the RASS regions used to estimate the Galactic foreground flux. The region excluded around the background cluster Abell 1553 is shown in yellow. Colorbar units are counts/s/arcmin$^2$.}\label{img}
\end{center}
\end{figure}

An image of the complete mosaic in the 0.7-2.0 keV energy band, corrected for instrumental background and vignetting, is shown in Figure \ref{img}. Point sources were identified and excluded from further analysis using the CIAO tool \texttt{wavdetect} employing a single wavelet radius of 1 arcmin, which is matched to the Suzaku half-power radius. In addition, we removed a 13$^\prime$ radius circle around the background cluster Abell 1553.
 
\section{Spectral modeling}

For each arm, we extracted spectra from partial annuli centered on the core of M87. We imposed a minimum of 1200 source counts per spectral extraction region, accounting for the instrumental and sky backgrounds estimated in Section \ref{cxb_default}. Due to the point-spread function (PSF) of Suzaku, the minimum width of each annulus was set to 3 arcmin. Along the southern arm, we only use the data out to a radius of 957~kpc, corresponding to the minimum surface brightness between M87 and the infalling halo of M49.

The spectral modelling was performed with the XSPEC spectral fitting package \citep[version 12.8.2,][]{arnaud1996}, employing the modified C-statistic estimator. The spectra were binned to a minimum of one count per channel and fitted in the 0.7--7.0 keV band. Unless noted otherwise, the ICM was modelled as a thermal plasma in collisional ionization equilibrium using the {\it vapec} code \citep{smith2001}. The metal abundances are reported with respect to the proto-solar units of \cite{lodders2003}. With respect to more recent Solar abundance measurements by \citet{Asplund2005}, the Si/Fe would be 14\% higher, S/Fe 25\% higher, and Mg/Fe 5\% lower than the values reported here.
The Galactic absorption column density, $n_H$, was fixed to the average value at the location of each respective Suzaku field based on the radio HI survey of \cite{kalberla2005}. 

\subsection{Background subtraction}\label{cxb_default}

To account for the instrumental and cosmic background, we follow the methods laid out in e.g. \cite{urban2014}. In short, the instrumental background was subtracted in the standard way, using the task \texttt{xisnxbgen} \citep{tawa2008}. The cosmic X-ray background model consisted of a power-law emission component that accounts for the unresolved population of point sources (CXB), one absorbed thermal plasma model for the Galactic halo (GH) emission, and an unabsorbed thermal plasma model for the Local Hot Bubble (LHB) \citep[see e.g.][]{kuntz2000}.

The average flux and spatial variation of the GH and LHB emission was determined using a set of 12 ROSAT All-Sky Survey (RASS) spectra obtained from circular regions with a radius of 1 degree (three regions beyond each Suzaku arm; see Figure \ref{img}). The metallicity of both components was assumed to be Solar. Six Suzaku pointings (the outermost three along the N and W directions, respectively) do not contain any significant ICM emission. We thus determined the parameters of the CXB model by fitting the spectra obtained from the full field of view of these observations in the 2-7 keV band. The best-fit model parameters and their estimated statistical and systematic uncertainties are summarised in Table \ref{tab_cxb}.

\begin{table}
\caption{Background model parameters. The first set of errors gives statistical uncertainties at 68\% confidence; the second set represents systematic uncertainties due to spatial variations.}
\begin{center}
\begin{tabular}{cccc}
\hline
\hline
 & kT (keV) & arm & XSPEC norm  \\
 & or $\Gamma$ & & per arcmin$^2$ ($\times 10^{-7}$) \\
\hline 
CXB & $1.53\pm0.04$ & all & $0.83\pm0.05\pm0.06 $ \\ 
\hline
 & & N & $1.31\pm0.10\pm0.22 $ \\
GH & $ 0.20\pm0.01$ & W & $1.14\pm0.10\pm0.12$ \\
 & & S & $2.20\pm0.15\pm0.66 $  \\
 & & E & $1.73\pm0.12\pm0.11$  \\
\hline
 & & N & $1.08\pm0.04\pm0.08 $  \\
LHB & $0.104\pm0.002$ & W & $1.35\pm0.05\pm0.05 $ \\
 & & S & $1.46\pm0.06\pm0.27 $ \\
 & & E & $1.41\pm0.05\pm0.11$  \\
\hline 
\end{tabular}
\end{center}
\label{tab_cxb}
\end{table}

We checked for solar wind charge-exchange (SWCX) emission contamination following the analysis of \citet{Fujimoto07}. Based on solar wind proton flux measurements from the WIND SWE (Solar Wind Experiment\footnote{http://web.mit.edu/afs/athena/org/s/space/www/wind.html}), in combination with estimates of the radius of the Earth's magneto-sheath along the Suzaku line of sight during our observations, we find that the level of SWCX contamination is expected to be negligible for all the pointings considered in the present analysis.

\section{Results}

\begin{figure*}
\begin{center}
\includegraphics[width=\columnwidth]{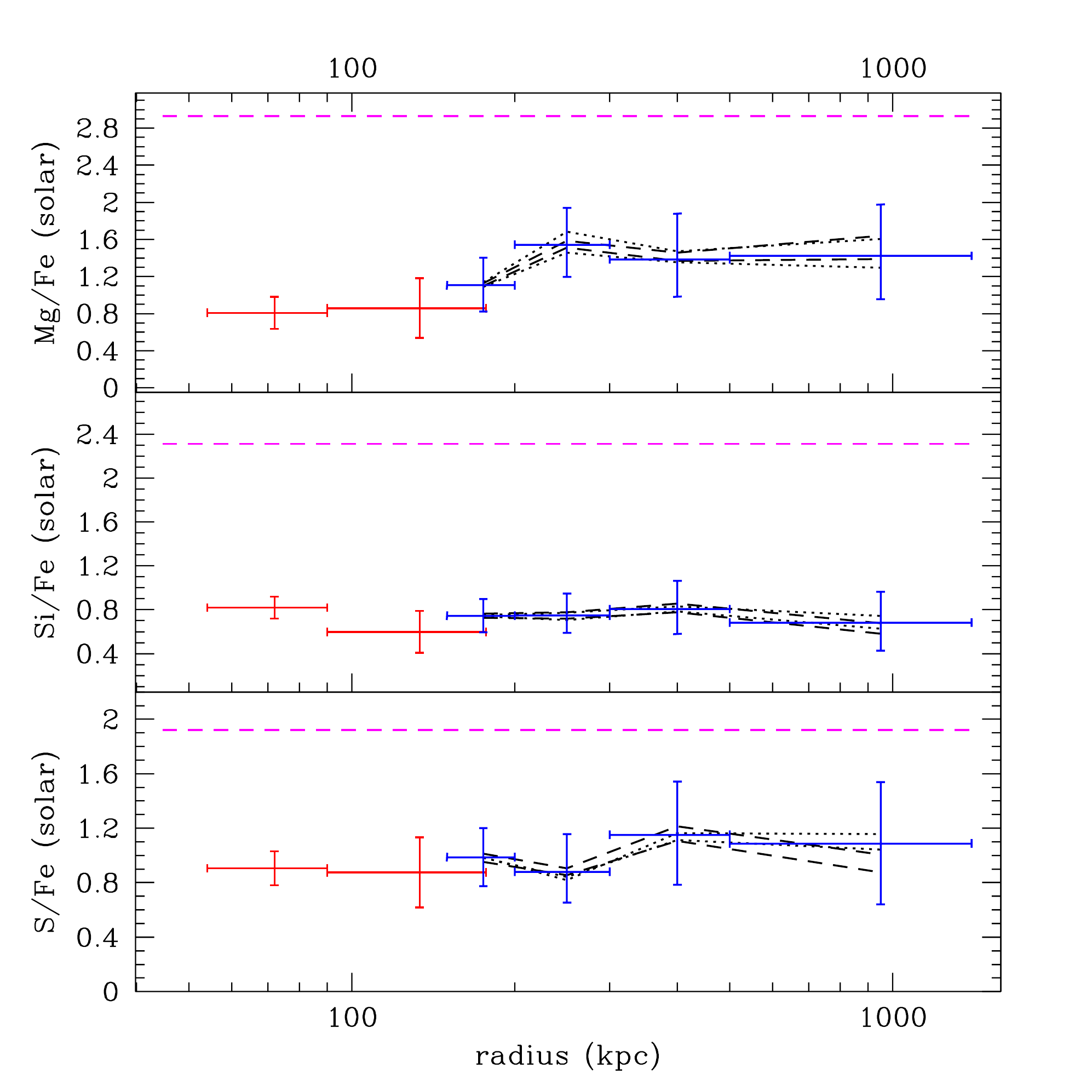}
\includegraphics[width=\columnwidth]{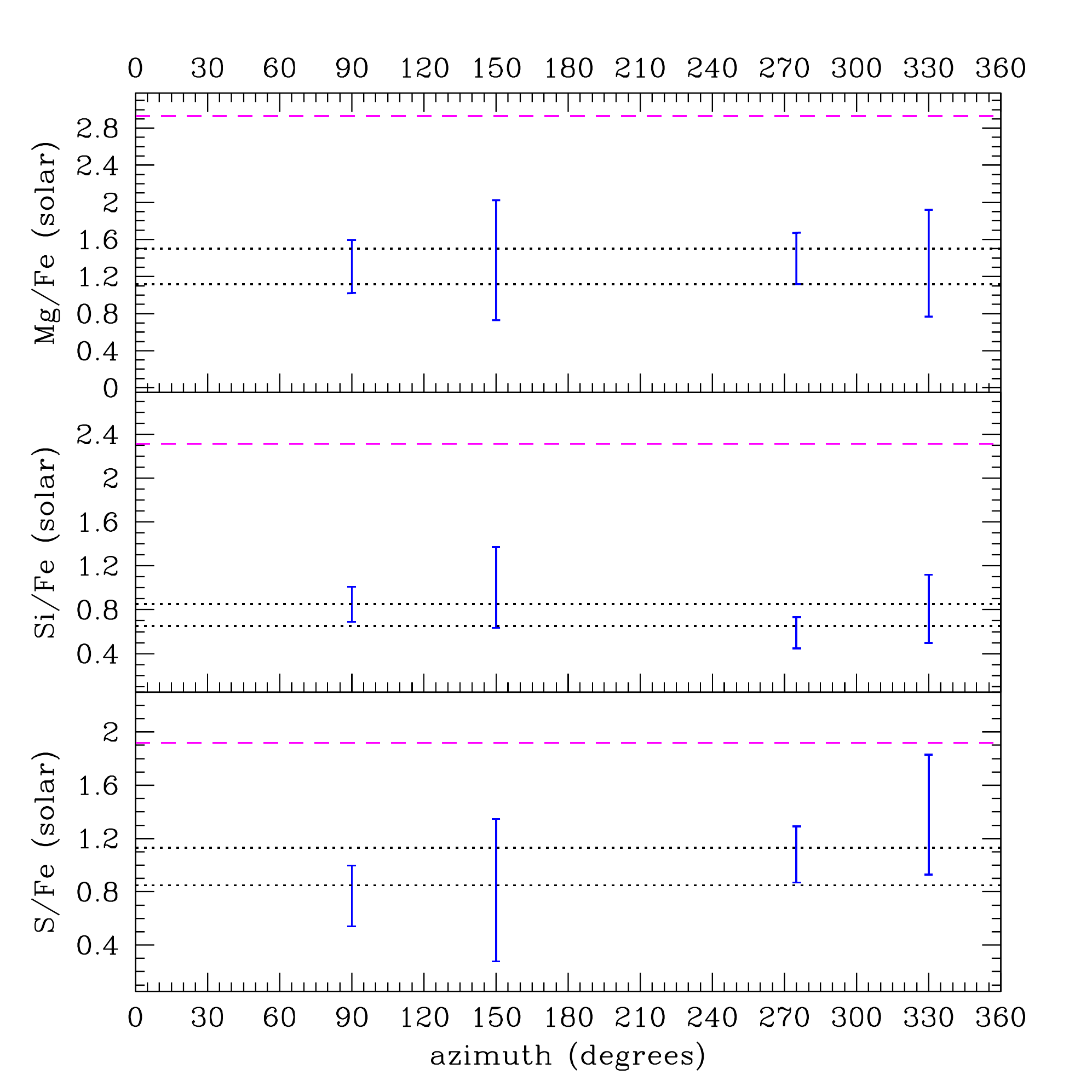}
\caption{{\it Left:} Abundance ratio profiles from the 2T \texttt{vapec} fit with $kT_{cool}=1.0$ keV are shown in blue. Red data points show published results from \cite{simionescu2010} for the N arm. Systematic uncertainties due to background fluctuations are shown with dotted lines for the GH and dashed lines for the CXB power-law. {\it Right:} Abundance ratios as a function of azimuth (measured counterclockwise of west). Dotted horizontal black lines show the $1\sigma$ confidence interval for the best-fit constant to the corresponding {\sl radial} profile. Magenta dashed lines show the expected values for enrichment by pure SNcc yields in both plots.}\label{mgsisfe}
\end{center}
\end{figure*}

In this Letter, we focus on measurements of the abundance of alpha-elements in the outskirts of the Virgo Cluster. A more detailed discussion regarding the thermodynamical properties of the ICM, and the total gas and metal masses compared to the stellar component, is the subject of a separate publication (Simionescu et al., in prep).

We divided the data into four radial bins, spanning 150--200 kpc, 200--300 kpc, 300--500 kpc, and from 500 kpc out to the edge of the detection limit. For each radial bin, we fit in parallel the spectra from all partial annuli at any azimuth that fall within the respective radial range. The temperature and spectrum normalization were left as free parameters for each partial annulus, while the Fe, Si, S, and Mg abundances were coupled to the same value. The abundances of all other elements between O and Ni were set to 0.2 Solar.
The results are summarised in Table \ref{tab}.

\begin{table}
\caption{Metal abundances as a function of radius for various spectral models. Errors are quoted at $\Delta C=1$.}
\begin{tabular}{ccccc}
\hline
\hline
 & 150--200 kpc & 200--300 kpc & 300--500 kpc & 500--1400 kpc \\
\hline 
 \multicolumn{5}{c}{vapec 1T} \\
 \hline
Mg & $0.16\pm0.06$ & $0.33\pm0.09$ & $0.15\pm0.09$ & $0.14\pm0.09$ \\
Si & $0.14\pm0.03$ & $0.17\pm0.05$ & $0.15\pm0.05$ & $0.10\pm0.05$ \\ 
S & $0.21\pm0.05$ & $0.22\pm0.07$ & $0.28\pm0.08$  & $0.21\pm0.08$ \\
Fe & $0.22\pm0.01$ & $0.28\pm0.02$ & $0.15\pm0.02$ & $0.17\pm0.02$ \\ 
\hline 
 \multicolumn{5}{c}{vapec 2T} \\
 \hline
Mg & $0.27\pm0.07$ & $0.50\pm0.11$ & $0.39\pm0.12$ & $0.32\pm0.11$ \\
Si & $0.18\pm0.04$ & $0.24\pm0.05$ & $0.23\pm0.07$ & $0.15\pm0.06$ \\ 
S & $0.24\pm0.05$ & $0.28\pm0.08$ & $0.33\pm0.10$ & $0.25\pm0.10$ \\
Fe & $0.24\pm0.01$ & $0.32\pm0.02$ & $0.28\pm0.02$ & $0.23\pm0.03$ \\ 
\hline 
 \multicolumn{5}{c}{spex 2T} \\
 \hline
Mg & $0.21\pm0.07$ & $0.48\pm0.10$ & $0.32\pm0.12$ & $0.45\pm0.12$ \\
Si & $0.20\pm0.04$ & $0.29\pm0.06$ & $0.25\pm0.05$ & $0.21\pm0.07$ \\ 
S & $0.24\pm0.05$ & $0.29\pm0.08$ & $0.33\pm0.11$ & $0.21\pm0.12$ \\
Fe & $0.29\pm0.02$ & $0.39\pm0.03$ & $0.32\pm0.04$ & $0.32\pm0.03$ \\ 
\hline 
\end{tabular}
\label{tab}
\end{table}

Biases in measuring the metallicities in the ICM, particularly that of Fe, can arise when X-ray emitting gas at several different temperatures contributes to the integrated spectrum, especially if the average temperature of the ICM is low \citep{buote2000}. We have thus refitted the data with a two-temperature model (hereafter 2T model or 2T fit), with the abundances of the two gas phases assumed to be the same, the normalisation of the cool component left free in the fit for each spectral extraction region separately, and the temperature of the cooler gas, $kT_{cool}$, successively fixed at 0.6, 0.8, 0.9, 1.0, 1.1, and 1.2 keV. The largest differences compared to the 1T fit are obtained for $kT_{cool}$=1 keV, and we show the results for this model in Table \ref{tab}. All metal abundances are systematically higher than those derived from a single-temperature fit; however, the metal abundance ratios are more robust, with Si/Fe and S/Fe largely consistent between the 1T and 2T models, while Mg/Fe is systematically higher by approximately $1\sigma$ for the two temperature fit in each radial bin compared to the single phase approximation.

Given that it is likely that multi-temperature structure is present in the ICM, and that more complex multi-phase models can not be constrained (and are unlikely to have a bigger impact than the difference between single-phase and 2T models), we have chosen to use the results from the 2T fit as reference values for later discussion. For this spectral model, in the radial bin beyond 500~kpc, we detect Mg with a statistical significance of 2.9$\sigma$, and Si and S each with a significance of 2.5$\sigma$; line emission from alpha-elements in the cluster outskirts is thus detected with a cumulative significance of $>4.5\sigma$. We have verified that the same level of significance is obtained when the statistical uncertainties are derived using a Markov chain Monte Carlo approach instead of the $\Delta C=1$ method adopted throughout the rest of this manuscript.

In the left panel of Fig. \ref{mgsisfe}, we show the metal abundance {\it ratio} profiles obtained from the 2T fit. All the alpha-element to Fe ratios are consistent with being constant as a function of radius all the way to the edge of the Virgo Cluster. The best-fit constants in the radial range between 150 and 1400 kpc are $0.75\pm0.10$, $0.99\pm0.14$, and $1.31\pm0.19$ for the Si/Fe, S/Fe and Mg/Fe ratios, respectively, therefore generally in good agreement with the Solar abundance pattern, with slightly sub-solar Si/Fe and super-solar Mg/Fe. Linear fits to the radial profiles of the abundance ratios yield a slope consistent with zero in all cases. Our results are in good agreement with the metallicity ratios measured within the central $0.5 r_{180}$ for four other galaxy groups \citep{Sasaki2014}. 

We have also tested for the {\sl azimuthal} dependence of the abundance ratios by fitting in parallel the spectra from all radii along a given Suzaku arm. The results are shown in the right panel of Figure \ref{mgsisfe}. We find that the measured abundance ratios are consistent with a flat distribution as a function of azimuth; along each arm, the measurements agree with the best-fit constant obtained from the corresponding radial profiles.

\subsection{Systematic uncertainties}\label{sect:sys}
 
To investigate the systematic uncertainties associated with the assumed plasma emission code, we have also refitted the data using the CIE plasma model implemented in \texttt{spex} \citep{kaastra1996} version 2.00.11. The results for a two-temperature fit with $kT_{cool}$ fixed at 1.0 keV are shown in Table \ref{tab}. 
The abundances of all the alpha-elements agree with those derived from the \texttt{vapec} 2T fit within the $1\sigma$ statistical error bars; however, due to differences in the Fe L-shell transitions between the two plasma codes, the Fe abundance is systematically higher when using \texttt{spex}. This only increases the significance with which a contribution from SN~Ia products would be required to explain the enrichment of the cluster outskirts (Section \ref{disc}).
 
We estimated the systematic uncertainties due to the background model by refitting the spectra with the GH and CXB fluxes scaled according to the expected fluctuations described in Table \ref{tab_cxb}, accounting for the $1/\sqrt{\Omega}$ dependence of the cosmic variance of unresolved point sources on the size of the extraction region $\Omega$. The results are shown in Figure \ref{mgsisfe}. The systematic uncertainties are in all cases smaller than the statistical error bars of the measurements. The contribution of the LHB and the associated systematic uncertainties are negligible in the energy band considered here (above 0.7~keV). The effects of a 5\% uncertainty on the instrumental background are comparable to those due to CXB fluctuations. Spatial variations of the Galactic column density do not influence the results significantly, such that consistent results are obtained assuming a uniform $n_H$ across the entire Suzaku mosaic. 

Note that, although the conclusions based on the Si abundance measurements provide a high statistical confidence, they may carry additional systematic uncertainties compared to S and Mg due to the complexity in calibrating the Si instrumental line as well as the effective area around the Si edge.

\section{Discussion and Conclusions}\label{disc}

We present the first measurements of the abundances of $\alpha$-elements extending out to beyond the virial radius of a galaxy cluster, specifically the nearby Virgo Cluster. Our main findings are summarised below.

\subsection{Pure SNcc enrichment ruled out in the cluster outskirts}

Assuming the SNcc yields of \cite{nomoto2006} for Solar metallicity of the progenitor and averaged over a Salpeter IMF, the expected Si/Fe, S/Fe and Mg/Fe ratios from enrichment of the ICM due solely to core-collapse supernova products are 2.31, 1.92, and 2.93 Solar, respectively (shown as dashed magenta lines in Fig \ref{mgsisfe}; these values increase if the metallicity of the progenitor is sub-solar). Therefore, the best-fit constant values describing our measured alpha-element to Fe abundance ratio profiles rule out pure SNcc enrichment beyond the cluster core with very high significance (15.6$\sigma$, 6.6$\sigma$, and 8.5$\sigma$ for Si/Fe of $0.75\pm0.10$, S/Fe of $0.99\pm0.14$, and Mg/Fe of $1.31\pm0.19$, respectively). Using only the data beyond half of the cluster's virial radius, pure SNcc enrichment is excluded at the 5.8$\sigma$ level based on the best-fit Si/Fe ratio, and 1.8$\sigma$ and 2.7$\sigma$ for the S/Fe and Mg/Fe measurements, respectively.

\subsection{Uniform chemical composition of the Universe over a wide range of scales}

Rather than following the chemical enrichment patterns expected from pure SNcc enrichment, as inferred from earlier work, the measured metal abundance ratios are largely consistent with the Solar value over the entire cluster volume probed by the present measurements (both as a function of radius and azimuth, from just outside the influence of M87 out to $1.3 r_{200}$). In addition, \citet{million2011} report detailed metal abundance ratios in the M87 halo using deep Chandra data, and find a Mg/Fe ratio also consistent with the Solar value from a radius of 7--40 kpc, while the Si/Fe and S/Fe ratios decrease from 1.2 Solar at a radius of 10~kpc to 1.0 Solar at 40~kpc. Generally, therefore, given the wide range of spatial scales probed, from the chemical content of stars like our Sun to the enrichment pattern on megaparsec scales, the chemical composition of the Universe appears to be remarkably uniform. 

This supports the conclusions based on previous Suzaku Key Project observations of the Perseus Cluster, which revealed a uniform Fe abundance distribution on large scales, indicating that most of the metal enrichment of the intergalactic medium occurred before the entropy in the ICM became stratified, probably during the period of maximal star formation and black hole activity at redshifts $z=2-3$ \citep{werner2013nat}. Such an early enrichment and mixing is likely responsible for producing the uniform metal abundance ratios observed today.

The average Fe abundance in the outer parts of the Virgo Cluster is $0.23\pm0.03$ Solar, marginally lower compared to $0.29\pm0.01$ in the Perseus Cluster (converted from \citealt{werner2013nat} to the Solar units adopted here). This difference may be due to uncertainties in the Fe-L line emissivities, since the \texttt{spex} plasma emission code suggests a higher Fe abundance that is in agreement with the Perseus Cluster measurement.

\subsection{The relative contribution of SN~Ia to the early enrichment of the intergalactic medium}

In order to provide a homogeneous chemical composition of the Universe, the rapid production of $\alpha$-elements by SNcc associated with the peak star formation activity at redshifts of 2--3 must have been complemented by a sufficient number of SN~Ia exploding during or shortly after this epoch. This is consistent with results from type Ia supernova delay-time distributions, which imply that a large fraction of such supernovae explode less than $\sim0.5$ Gyr after the formation of the progenitor binary system \citep{mannucci2006}. 

The ratio between the number of SN~Ia and the total number of supernovae contributing to the enrichment of the ICM, $R_{\%Ia}=N_{Ia}/(N_{Ia}+N_{CC})$, can be estimated by mixing model yields of SN~Ia and SNcc products in various percentages until the observed metal abundance ratios are reproduced. Using the same SNcc yields mentioned above, and assuming the SN~Ia yields based on the WDD2 model of \citet{iwamoto1999}, we obtain $R_{\%Ia}$ values of $37_{-8}^{+11}$\%, $18_{-5}^{+9}$\%, and $12^{+4}_{-2}$\%, corresponding to the measured Si/Fe, S/Fe, and Mg/Fe, respectively. Note however the additional possible systematic uncertainties associated with the Si/Fe determination (Section \ref{sect:sys}).

Although additional uncertainties due to different supernova yield models and assumptions about the IMF and initial metallicity of the progenitor stars remain \citep[see detailed discussion in e.g.][]{deplaa2007,deplaa2013}, the ratios presented here are generally consistent with those estimated for our own Galaxy \citep[15\%,][]{tsujimoto1995} or for the cluster cores \citep[about 20--30\%, e.g.][]{deplaa2007,simionescu2009b,simionescu2010}. 

Finally, note that the fractional contributions calculated here only reflect the ratio of SN~Ia to SNcc that enrich the ICM; if the process of injecting metals from the progenitors into the intergalactic medium has a different efficiency or timescale for different supernova types, the true $R_{\%Ia}$ may be different than reflected by the composition of the ICM.

\acknowledgments

We thank the anonymous referee for constructive comments, T. Sasaki and J. de Plaa for helpful discussion, and K. Nagayoshi for help with assessing the SWCX contamination. The work was supported in part by NASA grants NNX12AE05G and NNX13AI49G and by the US Department of Energy under contract number DE-AC02-76SF00515. YI is financially supported by a Grant-in-Aid for Japan Society for the Promotion of Science (JSPS) Fellows. 

{\it Facilities:} \facility{Suzaku}.

\bibliographystyle{apj}


\end{document}